\documentclass[aps,prd,preprint,groupedaddress]{revtex4-2}

\usepackage{dcolumn}   % needed for some tables
\usepackage{bm}        % for bold math symbols
\usepackage{amssymb}   % for math symbols
\usepackage{amsmath}   % for advanced math
\usepackage{hyperref}  % for hyperlinks

\newcommand{\GeV}{\text{GeV}}%{\text{\normalfont\AA}}
\newcommand{\cm}{\text{cm}}%{\text{\normalfont\AA}}

\begin{document}

\title{On quantum spreading of a localized stationary flow of high energy particles}

\author{N.F.~Shul'ga}
\email[]{shulga@kipt.kharkov.ua}
\affiliation{National Science Center "Kharkiv Institute of Physics and Technology". 1, Akademichna str. Kharkiv 61108 Ukraine,
Karazin Kharkiv National University. 4, Svobody sq. Kharkiv 61000 Ukraine}

\author{S.N.~Shulga}
\email[]{shulga.serge@gmail.com}
\affiliation{National Science Center "Kharkiv Institute of Physics and Technology". 1, Akademichna str. Kharkiv 61108 Ukraine,
Karazin Kharkiv National University. 4, Svobody sq. Kharkiv 61000 Ukraine}

\date{\today}

\begin{abstract}
The study addresses the quantum spreading of a localized stationary flow of high energy particles. Results demonstrate that as particle energy increases, the spreading speed of the particle wave packet diminishes rapidly. Concurrently, increasing the energies stabilizes the initially localized packet, preserving its transverse form in a vacuum over extended distances. This allows substantial simplifications when using various approximate methods to calculate the wave function in an external field.
\end{abstract}

%\begin{keyword}
%Quantum electrodynamics \sep high energy \sep wave packet spreading \sep geometrical optics \sep wave mechanics.
%\MSC[2010] 00-01\sep  99-00\\
%PACS numbers: 29.27.-a, 61.85.+p, 34.80.Pa, 61.05.J
%\end{keyword}

\maketitle

\section{\label{sec:Intro}Introduction}

When a fast charged particle interacts with a localized stationary external field, its total energy is conserved. 
The wave function of such a particle is determined by the solution to a stationary wave equation (Dirac equation for a particle with spin, and Klein-Gordon equation for a particle without spin). 
To determine the solutions to these equations, various approximate methods are employed.
These include the eikonal and semiclassical approximations, geometrical optics approximation, etc. (see \cite{ref1_AkhiezBerest_QED1965, ref2_BerLifPit_RelQuTh1968, ref3_AkhiezShulga_HighEn1996, ref4_KraOrl_GeomOptInhMed2011, ref5_Shulga_PLB2019} and references therein). 
When using these methods, the effects of wave packet spreading are typically neglected, simplifying the calculations significantly. 
In \cite{ref6_Blokh_HEPThEP1967} (see also \cite{ref3_AkhiezShulga_HighEn1996}), the case  of wave packet spreading over time for high-energy particles was examined. 
It was demonstrated that increasing the particle energies lead to the stabilization of wave packets.

In this study, we address the problem of spreading of a stationary flow of high energy particles, that follows from the solution of the stationary wave equation in vacuum. 
For simplicity, our analysis is based on the solution of the Klein-Gordon wave equation.

%%%%%%%%%%%%%%%%%%%%%%%%%%%%%%%%%%%%%%%%%%%%%%%%%%%%%%%%%%%%%%%%%%%%%%%%%%%%%%%%%%%%%%%%%%%%%%%%%%%%%%%%%%%%%%%%%%%%  
\section{\label{sec:QuantSpreading}Quantum spreading of a stationary flow of particles}
     
The stationary Klein-Gordon equation for a particle with the mass $m$ and energy $\varepsilon$ has the following form:
\begin{equation}
\label{eq_statKlGdEq}
\left[ \varepsilon^2-m^2-(-i\hbar\nabla)^2 \right] \phi(\bf{r})=0.
\end{equation}
We use here the system of units in which the speed of light is equal to one.
 
By factoring out a plane wave component from the wave function $\phi(\bf{r})$ we can express it in the form:
\begin{equation}
\label{eq_wf_plwf}
\phi({\bf{r}}) = e^{i{\bf{pr}}/\hbar} f(\bf{r}),
\end{equation}
where ${\bf p} = {\bf e}_z\sqrt{\varepsilon^2-m^2}$ is the particle momentum directed along the initial direction of motion that we denote as the $z$ axis (${\bf e}_z$ is the unity vector in that direction). Then the pre-exponential function $f(\bf{r})$ will, accordingly to \eqref{eq_statKlGdEq}, satisfy the equation
\begin{equation}
\label{eq_preexp}
\partial_{z}f = \frac{i\hbar}{2p}\left( \frac{\partial^2}{\partial z^2} + \bf{\nabla}_\bot^2 \right) f,
\end{equation}
where $\bf{\nabla}_\bot = {\partial}/{\partial{\bm\rho}}$ and ${\bm\rho}$ are the coordinates in the plane orthogonal to the $z$ axis.

In common case, the solution of \eqref{eq_preexp} can be found via Fourier transform of the function $f({\bm \rho},z)$:
\begin{equation}
\label{eq_ffourier}
f({\bm \rho},z) = \int{\frac{d^2q}{(2\pi)^2}e^{i{\bf qr}}f_{\bf q}(z)},
\end{equation}
with $f_{\bf q}(z)$ obeying the equation 
\begin{equation}
\label{eq_partfz}
\partial_{z}f_{\bf q}(z)=\frac{i\hbar}{2p}\left(
\frac{\partial^2}{\partial z^2}-{\bf q}^2
\right)f_{\bf q}(z).
\end{equation}

By passing from the function $f_{\bf q}(z)$ in \eqref{eq_partfz} to the function $\varphi_{\bf q}(z)$, which relates to $f_{\bf q}(z)$ as
\begin{equation}
\label{eq_fvarphi}
f_{\bf q}(z)=e^{-i\frac{\hbar q^2}{2p}z}\varphi_{\bf q}(z),
\end{equation}
we obtain the subsequent equation for $\varphi_{\bf q}(z)$:
\begin{equation}
\label{eq_varphi_q_z}
\partial_z\varphi_{\bf q}(z) = \frac{i\hbar}{2p}\left[
\frac{\partial^2}{\partial z^2}\varphi_{\bf q} - \frac{i\hbar q^2}{p}\partial_z\varphi_{\bf q} - 
\left(
\frac{\hbar q^2}{2p}
\right)^2 \varphi_{\bf q}
\right].
\end{equation}

Let us consider the solution of \eqref{eq_varphi_q_z} that corresponds to the initial state 

\begin{equation}
\label{eq_phi_rho_z_0}
\left.\phi({\bm\rho},z)\right|_{z=0} = 
\frac{1}{\sqrt{\pi}a}e^{-\rho^2/2a^2}
\end{equation}
($a$ is the packet width at $z=0$), and to the normalization condition

\begin{equation}
\label{eq_phi_rho_z_norm}
\int{d^2\rho \left|\phi({\bm\rho},z)\right|^2}=1.
\end{equation}

At $z=0$, for the function $\varphi_{\bf q}(z)$ we therefore have 
\begin{equation}
\label{eq_varphi_q_z0}
\left. {\varphi_{\bf q}} \right|_{z=0} = 2\sqrt{\pi}a e^{-q^2a^2/2}.
\end{equation}

The right-hand side of equation \eqref{eq_varphi_q_z} contains terms proportional to the inverse powers of the momentum $p$. Thus, the solution to this equation can be sought as a series in inverse powers of $p$.
The first terms of this expansion, which satisfy the initial condition \eqref{eq_phi_rho_z_norm}, take the following form:
\begin{equation}
\label{eq_varphi_q_z_develp}
\varphi_{\bf q}(z) = 2\sqrt{\pi}a \left[
	1-\frac{i\hbar}{2p}\left(
		\frac{\hbar q^2}{2p}
	\right)^2 z
\right]
e^{-q^2a^2/2}.
\end{equation}

When we substitute \eqref{eq_varphi_q_z_develp} into \eqref{eq_fvarphi} and \eqref{eq_ffourier}, it becomes evident that for $z \lesssim pa^2/\hbar$ the primary contribution to the integral in \eqref{eq_ffourier} comes from values of $q \sim \sqrt{2}/a$.
In this scenario, we derive the following estimate for the term in \eqref{eq_varphi_q_z_develp} that is proportional to $p^{-2}$:
\begin{equation*}
\label{}
\frac{\hbar z}{2p}\left(
\frac{\hbar q^2}{2p}
\right)^2 \propto
\left(
\frac{\hbar}{pa}
\right)^2
\ll 1.
\end{equation*}
By neglecting the term in \eqref{eq_varphi_q_z_develp}, we obtain the following expression for $\phi({\bm\rho},z)$: 
\begin{equation}
\label{eq_phi_rho_z}
\begin{split}
\phi({\bm\rho},z) &= 
\frac{1}{\sqrt{\pi}a}
\cdot
\frac{1}{ 1+i{\frac{\hbar z}{pa^2}} }\times \\
&\times\text{exp}
\left[
i\left(
\frac{pz}{\hbar} + \alpha({\bm\rho},z)
\right) - 
\frac{\rho^2/2a^2}{1+\left(\frac{\hbar z}{pa^2}\right)^2}
\right]
\end{split}
\end{equation}
where
\begin{equation}
\label{}
\alpha({\bm\rho},z) = \frac{\hbar z}{pa^2}\cdot\frac{\rho^2/2a^2}{1+(\hbar z/pa^2)^2}.
\end{equation}

It is obvious that the wave function \eqref{eq_phi_rho_z} fulfills the normalization condition \eqref{eq_phi_rho_z_norm}. Concurrently, for $z \lesssim pa/\hbar$, the average value of the square of the packet width over the length $z$ is governed by the relation:
\begin{equation}
\label{eq_rhosq_mean}
\overline{\rho^2(z)} = \frac{a^2}{1+(\hbar z/pa^2)^2}.
\end{equation}
%
%

%%%%%%%%%%%%%%%%%%%%%%%%%%%%%%%%%%%%%%%%%%%%%%%%%%%%%%%%%%%%%%%%%%%%%%%%%%%%%%%%%%%%%%%%%%%%%%%%%%%%%%%%%%%%%%%%%%%%%%%%%%%%%%%%%%%%%%%%%%%%

\section{\label{sec:Conclusioni}Conclusions}

The formula \eqref{eq_phi_rho_z} suggests that an initially localized at $z=0$ stationary flow spreads in the transversal direction at $z>0$. The rate of this spreading diminishes rapidly with increasing particle energy. 
The transversal spreading of a flow increases to a value $\Delta\rho\sim a$ at distances

\begin{equation}
\label{eq_zsim}
z \sim pa^2/\hbar,
\end{equation}
which are considerably greater than $a$.
So, for electrons with the energy $\varepsilon=1\;\GeV$ and $a\sim 10^{-8}\,\cm$ this is reached in distances $z$ of the order of $10^{-2}\,\cm$. This effect arises from the quantum wave nature of the process. We see that, with the increase of the particle energy a stabilization of the initially localized particle flow takes place. 
In this case, the flow under study can be viewed as a set of rays localized within the region $\Delta\rho\sim a$, each characterized by the phase $\varphi=pz/\hbar+\alpha({\bm\rho},z)$.
For distances $z\ll pa^2/\hbar$, the effect of quantum spreading of the packet can be neglected according to \eqref{eq_phi_rho_z}.
In this scenario, the phase $\alpha({\bm\rho},z)$ is also small, which is associated with the process of packet spreading.
Consequently, for distances $z\ll pa^2/\hbar$ where the particle interacts with an external localized field (e.g., field of atom, ensemble of atoms, etc.), the wave function of the particle will be shaped by both the motion dynamics of the collective rays within this field and the corresponding phases.
It is worth noting that an analogical stabilization effect of wave packets at high energies was also predicted in \cite{ref6_Blokh_HEPThEP1967} during the analysis of the time-based spreading of a localized three-dimensional wave packet. 
Furthermore, this study also revealed that the wave packet spreads more slowly in the longitudinal direction compared to the transversal one. 

The results obtained above relate to the analysis of the problem of the quantum spreading of a stationary particle flow. These insights provide a basis for estimating the significance of this process when employing various approximate methods to determine wave equation solutions in high-energy physics.
 
%%%%%%%%%%%%%%%%%%%%%%%%%%%%%%%%%%%%%%%%%%%%%%%%%%%%%%%%%%%%%%%%%%%%%%%%%%%%%%%%%%%%%%%%%%%%%%%%%%%%%%%%%%%%%%%%%%%%%%%%%%%%%%%%%%%%%%%%
 
\begin{acknowledgments}
This research was partially supported by the National Academy of Sciences of Ukraine (project No. 0121U111556), it was partially conducted in the scope of the IDEATE International Research Project. Part of the research was supported by the Francophone University Association.
\end{acknowledgments}

\appendix

\section{Appendix Title}
Appendix content.

%\bibliography{yourbibfile}  % your .bib file here

\end{document}